\documentclass[12pt]{article}

\usepackage{amsmath,amsfonts,amssymb}
\usepackage{graphicx, psfrag}
\newcommand{\be}{\begin{equation}}
\newcommand{\ee}{\end{equation}}
\newcommand{\bea}{\begin{eqnarray}}
\newcommand{\eea}{\end{eqnarray}}
\newcommand{\ba}{\begin{array}}
\newcommand{\ea}{\end{array}}

\begin{document}

\begin{titlepage}
\noindent
{\tt IITM/PH/TH/2006/16}\hfill
{\tt hep-th/0612302}\\
\hfill December 2006 \\

\vspace{20pt}

\begin{center}
\textbf{\Large Toric K\"ahler metrics and  $AdS_5$ in  \\ \vspace{10pt} ring-like co-ordinates}\\

\vspace{20pt}

Bobby S. Acharya$^a$\footnote{bacharya@cern.ch}, 
Suresh Govindarajan$^b$\footnote{suresh@physics.iitm.ac.in} and  
Chethan N. Gowdigere$^a$\footnote{cgowdige@ictp.it}

\vspace{10pt} $^a$\textit{Abdus Salam ICTP, Strada Costiera 11,}\\
\textit{34014 Trieste, ITALY.}\\
\vspace{10pt}$^b$\textit{Department of Physics,}\\
\textit{Indian Institute of Technology Madras, Chennai 600036, INDIA.}\\

\end{center}
\vspace{10pt}

\begin{abstract}

Stationary, supersymmetric supergravity solutions
in five dimensions have K\"ahler metrics 
on the four-manifold orthogonal to the 
orbits of a time-like Killing vector. 
We show that an explicit
class of toric K\"ahler metrics provide a unified framework in which to
describe both the asymptotically flat and asymptotically $AdS$ solutions. 
The Darboux co-ordinates used for the 
local description turn out to be ``ring-like.'' 
We conclude with an Ansatz for studying the existence of 
supersymmetric black rings in
$AdS$.

\end{abstract}

\end{titlepage}

\section{Introduction and Summary}

Black holes can have non-spherical horizons in more than four
spacetime dimensions. In five dimensions, one encounters 
the first example of a non-spherical horizon, $S^2 \times S^1$, a 
black ring\footnote{For a review, see \cite{Emparan:2006mm}.}.  Emparan and 
Reall constructed the first black ring solution as a solution to 
vacuum gravity \cite{Emparan:2001wn}. Solutions in five dimensions can have two 
independent rotation parameters, call them $J_{\psi}$ and 
$J_{\phi}$. The Emparan-Reall ring rotates only along the 
direction of the ring, $J_{\phi} = 0$. Subsequently, black-ring 
solutions with $J_{\psi} = 0$ but $J_{\phi} \neq 0$ were discovered \cite{Figueras:2005zp}, again
in vacuum gravity. 
A black ring 
solution to Einstein-Maxwell theory was constructed in \cite{Emparan:2004wy}.

Asymptotically flat supersymmetric black rings were constructed 
first in Elvang et al \cite{Elvang:2004rt}, though they had been conjectured to 
exist earlier in \cite{Bena:2004wt}.  We will refer to this solution as 
the \emph{flat susic ring.} The flat susic ring has both the 
rotation parameters turned on, $J_{\psi} \neq 0, J_{\phi} \neq 
0$. This solution fits into the classification scheme for 
supersymmetric solutions to five dimensional minimal 
supergravity, discovered by Gauntlett et. al. \cite{Gauntlett:2002nw}. According 
to this classification, there are two classes of solutions, 
corresponding to whether the Killing spinor bilinear 
$\overline{\epsilon} \, \gamma^\mu\, \epsilon$, which is always a 
Killing vector, is time-like or null. In this paper we will only 
be interested in the time-like class; choosing the time 
co-ordinate, $\tau$, along the orbits of the time-like Killing 
vector, the metric takes the following stationary form:
\begin{equation}\label{01}
ds^2 = - f^2 \,(d\tau + \omega)^2 + \frac{1}{f} \,ds^2_{M_4}
\end{equation}
where $f$ is a function, $\omega$ is a one-form and $ds^2_{M_4}$ 
is a four-metric on the space transverse to the orbits of the 
Killing vector. $M_4$ will be referred to as the 
\emph{base-space}. It was shown in \cite{Gauntlett:2002nw} that supersymmetry 
requires $f$ to be a function on $M_4$, $\omega$ to be a one-form 
on $M_4$ and that the base-space be a hyper-K\"ahler manifold. 
These quantities must also satisfy the following p.d.e's.
\begin{equation}\label{02}
d \,G\,^+ = 0, \qquad \Delta \left( 1/f \right) = \frac{4}{9}\, |G\,^+|^2,
\end{equation}
where 
\begin{equation}\label{021}
G\,^\pm = \frac{f}{2} \,( d\,\omega \,\pm\, \ast \,d\,\omega )
\end{equation}
and $\Delta$ is the Laplacian for functions on the base-space. 
For the susic flat ring solution of \cite{Elvang:2004rt}, the base-space is 
just flat-space; however, Euclidean coordinates are not so useful here and
one writes the flat-space metric in 
special co-ordinates that will 
be referred to as \emph{C-metric} co-ordinates. In these 
coordinates, the above 
equations become simple to solve\cite{Bena:2004de}. There are other 
supersymmetric black 
ring solutions which involve more non-trivial hyper-K\"ahler 
base-spaces: the Gauntlett-Gutowski rings \cite{Gauntlett:2004wh,Gauntlett:2004qy} that 
have a Gibbons-Hawking base-space.

An interesting question is the existence of supersymmetric black ring 
solutions in $AdS$ space, i.e. ring solutions to gauged 
supergravity theories. The theory of supersymmetric solutions to 
gauged supergravity was first worked out by Gauntlett and 
Gutowski in \cite{Gauntlett:2003fk}, henceforth referred to as the 
Gauntlett-Gutowski theory, for the minimal case and subsequently 
for the $U(1)^3$ case in \cite{Gutowski:2004ez}. Again there is a time-like 
class of solutions, whose metric takes the form \eqref{01} and 
as in the asymptotically flat case, $f$ is a function on the 
base-space and $\omega$ is a one-form on the base-space. But now 
supersymmetry requires, both in the minimal and $U(1)^3$ cases, 
that the base-space $M_4$ be a K\"ahler manifold.

We will import a class of $U(1)^2$ invariant K\"ahler metrics from the 
mathematics literature \cite{6}. The metrics are given in Darboux 
or symplectic co-ordinates; in these co-ordinates {\it all the metrics 
in the class have the same K\"ahler form}, whilst the complex structures
can be different.
We find in this class two important metrics: 
\textbf{1.} the base-space metric for the susic flat ring and 
\textbf{2.} the base-space metric for $AdS_5$, i.e. the Bergmann 
metric. The parameter that deforms the first to the second is 
precisely the cosmological constant.
\begin{itemize}
\item[\textbf{1.}] For the flat-space metric, we find that the 
Darboux 
co-ordinates are ``ring-like'': the co-ordinate ranges of the 
Darboux co-ordinates in the co-ordinate directions orthogonal to 
the $U(1)^2$ directions are identical to the corresponding 
C-metric co-ordinates and the flat-space metric in Darboux 
co-ordinates is isometric to the flat-space metric in C-metric 
co-ordinates.  We give the explicit co-ordinate transformations.

\item[\textbf{2.}] We show that a certain metric in the class of 
metrics \cite{6} is the Bergmann metric. The co-ordinate region 
in which this form of the Bergmann metric is defined is exactly 
the same as the co-ordinate region in which the flat-space metric 
is defined. We thus obtain the Bergmann metric and hence $AdS_5$
in ring-like  co-ordinates. We give the explicit co-ordinate transformation between the polar co-ordinates in which the Bergmann metric is usually expressed and the ``ring-like'' Darboux co-ordinates.  
\end{itemize}

Since in the Darboux co-ordinates, the K\"ahler form takes such a 
simple form, the underlying K\"ahler structure of susic flat 
solutions becomes prominent when we express them in these 
co-ordinates.  We find that the one-form $\omega$ of the susic 
flat ring takes the following simple form:
\begin{equation}\label{022}
\omega = \rho~ d\psi + k ~\Omega^{(1)},
\end{equation}
where $\Omega^{(1)}$ is a one-form whose exterior derivative is 
the K\"ahler form, $\psi$ is the co-ordinate along the ring and 
$\rho$, $k$ are functions. Apart from the susic flat ring, there 
are other $U(1)^2$ invariant asymptotically flat supersymmetric 
solutions known. These describe a supersymmetric black hole 
placed in the centre of a susic flat ring \cite{Gauntlett:2004wh,Gauntlett:2004qy,Bena:2004de,Bena:2005zy}. 
We describe these solutions in ring-like coordinates and show that the 
one-form $\omega$ also takes the form \eqref{022}.

Recent work \cite{Kunduri:2006uh} suggests that supersymmetric black rings do
not exist in $AdS_5$. Our Ansatz provides a framework for addressing
this interesting question further.

This paper is organized as follows. In section two, we reproduce 
all the necessary definitions, theorems  from the math 
papers 
that we will need for our work. In section three, we consider 
flat-space in Darboux co-ordinates, describe in what sense they are 
``ring-like'' and describe the relation between Darboux 
and C-metric coordinates.  In section four, we elaborate on the properties of 
a certain $U(1)^2$ invariant metric that allows us to conclude 
that it is the Bergmann metric, and obtain $AdS_5$ in 
``ring-like'' co-ordinates. In section five, we co-ordinate 
transform various flat supersymmetric solutions from C-metric 
co-ordinates to the Darboux co-ordinates to find the form 
\eqref{022} for the one-form $\omega$. In section six, we make 
Ans\"atze for the metric and the one-form of the $AdS$ ring and 
conclude with directions for future work.

\section{Toric K\"ahler metrics in Darboux co-ordinates}

The expectation that a supersymmetric $AdS$ ring should have a 
$U(1)^2$ isometry, similar to the susic flat ring, leads us to 
look for K\"ahler metrics preserving a $U(1)^2$ isometry. Such 
metrics have appeared before in both the math and physics 
literature and are referred to as \emph{ toric K\"ahler} metrics.  
More precisely, a toric K\"ahler metric is a K\"ahler 
metric admitting commuting holomorphic Killing vector fields 
which are independent.  There exists a description of such metrics 
in local co-ordinates in the symplectic geometry literature. One 
thinks of the K\"ahler form as a symplectic form; in symplectic 
geometry, one uses local co-ordinates, called symplectic or 
Darboux co-ordinates, in which the symplectic form takes a 
standard form while the metric and the complex structure are 
described by non-trivial tensors.  The study of K\"ahler metrics 
in symplectic co-ordinates is attributed to the mathematicians, 
Guillemin \cite{8} and Abreu \cite{9}. Symplectic co-ordinates 
also appear naturally in the context of the gauged linear sigma 
model \cite{Witten:1993yc}. Here, we reproduce the relevant 
proposition from \cite{6}. 
\vspace{15pt}
\newline \textbf{Proposition}. \emph{Let $G_{ij}$ be a positive 
definite $ 2 \times 2$ symmetric matrix of functions of 
2-variables $x_1$ and $x_2$ with inverse $G^{ij}$. Then the 
metric
\begin{equation}\label{21}
\sum_{i,j} \left( G_{ij} dx_i dx_j + G^{ij} dt_i dt_j \right)
\end{equation}
is almost-K\"ahler with K\"ahler form
\begin{equation} \label{22} 
\Omega = dx_1 \wedge dt_1 + dx_2 \wedge dt_2 
\end{equation} 
and has independent hamiltonian Killing vector fields 
$\partial/\partial t_1$, $\partial/\partial t_2$ with 
Poisson-commuting momentum maps $x_1$ and $x_2$. Any 
almost-K\"ahler structure with such a pair of Killing vector 
fields is of this form (where the $t_i$ are locally defined up to 
an additive constant), and is K\"ahler if and only if $G_{ij}$ is 
the Hessian of a function of $x_1$ and $x_2$. }
\vspace{15pt}

The function of two variables is known as the \emph{symplectic 
potential}. The K\"ahler metric that appears in the above 
proposition is still quite generic, being specified by an 
arbitrary function of two variables. Note that the two-form
$\Omega$ takes a simple form in symplectic co-ordinates while
the metric (and complex structure) are non-trivial. The 
authors of \cite{6} 
define a sub-class of toric K\"ahler metrics viz. 
\emph{ortho-toric K\"ahler} metrics.
\vspace{15pt}
\newline \textbf{Definition}. \emph{A K\"ahler metric is 
\emph{ortho-toric} if it admits two independent hamiltonian 
Killing vector fields with Poisson-commuting moment maps $(\xi + 
\eta)$ and $ \xi \,\eta $ such that $d\xi$ and $d\eta$ are 
orthogonal. }
\vspace{15pt}

In other words, an ortho-toric K\"ahler metric is a toric 
K\"ahler metric, which when expressed in the $\xi$-$\eta$ 
co-ordinates does not contain cross-terms ($g\,^{\xi \eta} = 0$.)
\begin{equation} \label{23}
x_1 = \xi + \eta, \qquad x_2 = \xi \,\eta. 
\end{equation}
One of the virtues of the $\xi$-$\eta$ co-ordinates is that there 
is a symmetry under the exchange $\xi \leftrightarrow \eta$, 
which simplifies computations and allows for compact expressions. 
This feature is shared by the $x$-$y$ part of the C-metric 
coordinates, about which we will elaborate later. We will 
henceforth refer to the $\lbrace \xi, t, \eta, z \rbrace$ 
co-ordinates as the \emph{Darboux coordinates}, even if that 
term strictly should mean the $\lbrace x_1, t, x_2, z \rbrace$ 
co-ordinates. We will now quote the following proposition from 
\cite{8}, which provides a local form of ortho-toric K\"ahler 
metrics in the $\xi$-$\eta$ coordinates.
\vspace{15pt}
\newline \textbf{Proposition.} \emph{ The almost-Hermitian 
structure $ ( g, J, \Omega )$ defined by
\begin{equation}\label{24}
g = (\xi - \eta) \left( \frac{d\xi^2}{F(\xi)} - \frac{d\eta^2}{G(\eta)} \right)  + \frac{1}{\xi - \eta} \left( F(\xi) (dt + \eta \,dz)^2 - G(\eta) ( dt + \xi \,dz)^2 \right),
\end{equation}
\begin{align}\label{25}
J d\xi & = \frac{F(\xi)}{\xi - \eta} ( dt + \eta \,dz),& J dt & = -\frac{\xi\, d\xi}{F(\xi)} - \frac{ \eta \,d\eta}{G(\eta)}, \nonumber \\
J d\eta & = \frac{G(\eta)}{\eta - \xi} ( dt + \xi \,dz ),& J dz & = \frac{ d\xi}{F(\xi)} + \frac{d\eta}{G(\eta)},
\end{align}
\begin{equation}\label{26}
\Omega = d(\xi+\eta) \wedge dt + d(\xi \,\eta) \wedge dz
\end{equation} 
is an ortho-toric K\"ahler structure for any functions F,G of one 
variable. Every ortho-toric K\"ahler structure is of this form, 
where $t$, $z$ are locally defined up to an additive constant. }
\vspace{15pt}

The following one-form $\Omega^{(1)}$ whose exterior derivative 
gives the K\"ahler form \eqref{26} will appear later.
\begin{equation}\label{281}
d ~\Omega^{(1)} = \Omega, \qquad \Omega^{(1)} ~\equiv~ (\xi + \eta) \,dt + (1 + \xi \eta ) \,dz
\end{equation} 
It is worth re-emphasizing that all the K\"ahler metrics in 
\eqref{24} and \eqref{21} have the same $\Omega$ and 
$\Omega^{(1)}$ as given in \eqref{26} and \eqref{281}.

We will refer to the metrics \eqref{24} as the \emph{ACG} 
metrics. The authors of \cite{6} derive their motivation to 
consider co-ordinates \eqref{23} and to focus on the ortho-toric 
sub-class of toric K\"ahler metrics from their study of what are 
called \emph{weakly self-dual} K\"ahler metrics. A weakly 
self-dual K\" ahler metric is a K\"ahler metric whose 
anti-self-dual part of the Weyl tensor is harmonic, (after having 
fixed the underlying orientation to make the K\"ahler form 
self-dual.) A weakly self-dual K\"ahler metric has many important 
properties, one of which is that two particular functions built 
out of the metric are Poisson-commuting holomorphic potentials (a 
holomorphic potential is the moment map of a hamiltonian Killing 
vector field.) Poisson-commutation ensures that the corresponding 
hamiltonian Killing vector fields commute, though they need not 
be independent. The two functions in question are the normalized 
Ricci scalar,\footnote{This turns out to be $\frac{1}{6}$ of the 
Ricci scalar. The numerical factor is important while defining 
the co-ordinate transformation \eqref{27}.} $s$, and 
the pfaffian of the normalized Ricci form,\footnote{ The 
normalized Ricci form is the two-form $\frac{1}{2} \rho_0 + 
\frac{s}{4} \Omega$, where $\rho_0$ is the traceless Ricci form; 
and the pfaffian of a two-form $\psi$, $\textrm{pf}(\psi)$, is 
the 
function that multiplies the volume form to give $\psi \wedge 
\psi$, i.e. $\psi \wedge \psi = \textrm{pf}(\psi)\ vol$.} $p$. 
Furthermore, 
when expressed in coordinates related to the holomorphic 
potentials,
\begin{equation}\label{27}
s = \xi + \eta, \qquad p = \xi \,\eta,
\end{equation}
it turns out that a weakly self-dual K\"ahler metric does not 
contain cross-terms, i.e. $g\,^{\xi \eta} = 0$.

One can consider a ortho-toric weakly-self dual K\"ahler metric, 
i.e. a weakly self-dual K\"ahler metric for which the two Killing 
vector fields (whose moment maps are $s$ and $p$) are independent 
and commuting. From the fact that they are ortho-toric, they 
should take the form \eqref{24} and the weak self-duality 
condition should pick out specific forms for $F(\xi)$ and 
$G(\eta)$. It turns out that for the following choices of 
$F(\xi)$ and $G(\eta)$,
\begin{equation}\label{28}
F(x) = k\, x^4 + l \,x^3 + A \,x^2 + B \,x + C_1, \qquad G(x) = k\, x^4 + l\, x^3 + A \,x^2 + B\, x + C_2,
\end{equation}
where $k, l, A, B, C_1, C_2$ are constants, the ortho-toric 
metric \eqref{24} is weakly self-dual \cite{6}. 

The two cases of 
interest to us are when we have a quadratic polynomial for which 
the weakly self-dual metric is just flat space and when we have a 
cubic polynomial for which the weakly self-dual metric is a 
K\"ahler Einstein space. The former is the base-space for the 
full solution of a supersymmetric flat ring (more on this in 
section three) and the latter is the base-space for $AdS_5$ (more 
on this in section four). Therefore we see that these coordinates
will be useful in the description of several known, important examples.

\section{Flat space metric in Darboux co-ordinates}

In this section, we will first gather various facts about the 
C-metric co-ordinates $ \lbrace x, \phi, y, \psi \rbrace$, which 
will be relevant for the discussion on and comparison with the 
flat space metric in Darboux co-ordinates in the later 
subsection.

\subsection{Flat space metric in C-metric co-ordinates}

\begin{figure}[t]
\centering
\includegraphics[scale=0.45]{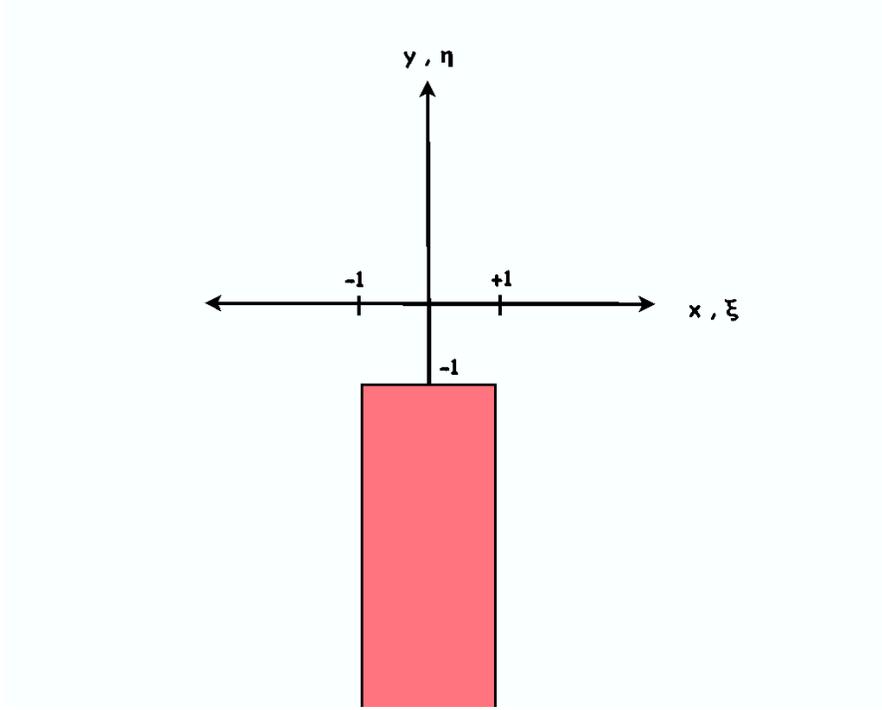}
\caption{The region in the $x$-$y$ and $\xi$-$\eta$ planes, where 
the metrics \eqref{46}, \eqref{411} and \eqref{31} are defined. 
Thus the supersymmetric flat ring and $AdS_5$ are defined in this 
co-ordinate region. }
\label{fig1}
\end{figure}

Following is the flat-space metric  in C-metric co-ordinates,
\begin{equation}\label{46}
ds^2 = \frac{1}{(x-y)^2} \left[ \frac{dx^2}{1-x^2} +(1-x^2) d\phi^2 + \frac{dy^2}{y^2-1}  +(y^2-1) d\psi^2 \right].
\end{equation}
The co-ordinate ranges for the $x$ and $y$ co-ordinates can be 
inferred from \eqref{46} by requiring that the metric be positive 
definite:
\begin{equation}\label{47}
-1 \leq x \leq 1, \qquad - \infty < y \leq -1.
\end{equation}
The metric \eqref{46} is thus defined in the region of the $x-y$ 
plane, shown in figure \ref{fig1}. The following nice physical 
interpretation of the C-metric co-ordinates was given in 
\cite{Emparan:2006mm}. In flat space, given in the usual polar co-ordinates,
\begin{equation}\label{41}
ds^2 = dr_1^2 + r_1^2 d\phi^2 + dr_2^2 + r_2^2 d\psi^2,
\end{equation}
a circular string of unit radius stretched along the $\psi$ 
direction, i.e. at $r_1 = 0, r_2 = \psi$, acting as an electric 
source for the three-form field strength $H = dB$, produces a 
field with only the following non-zero two-form potential:
\begin{equation}\label{42}
B_{t\psi} = - \frac{1}{2} \left(1 -\frac{1+r_1^2+r_2^2}{\Sigma} \right),
\end{equation}
where
\begin{equation}\label{43}
\Sigma = \sqrt{( 1 + r_1^2 + r_2^2\, )^2 - 4\,  r_2^2}
\end{equation}
and the two-form field strength dual to $H$ has a one-form 
potential ($\ast H = F = d A$) whose only non-zero component is
\begin{equation}\label{44}
A_\phi = - \frac{1}{2} \left( 1 + \frac{1 - r_1^2  - r_2^2}{\Sigma} \right).
\end{equation}
If one were to define new co-ordinates 
\begin{equation}\label{45}
x=\frac{1 -r_1^2-r_2^2}{\Sigma}, \qquad y=-\frac{1 +r_1^2+r_2^2}{\Sigma},
\end{equation}
then constant $A_\phi$ surfaces would be constant $x$ surfaces 
and constant $B_{t \psi}$ surfaces would be constant $y$ 
surfaces. These new co-ordinates $ \lbrace x, \phi, y, \psi 
\rbrace$ are nothing but the C-metric co-ordinates. The metrics 
\eqref{41} and \eqref{46} are isometric which can be explicitly 
verified with \eqref{45} and it's inverse:
\begin{equation}\label{49}
r_1 =  ~\frac{\sqrt{1-x^2}}{x-y}, \qquad r_2 =  ~\frac{\sqrt{y^2-1}}{x-y}.
\end{equation}
We thus see that the C-metric co-ordinates are ``ring 
co-ordinates,'' adapted to describe the fields set up by a 
charged ring source.

\subsection{ACG-quadratic metric}

Choosing the same quadratic polynomial for $F(\xi)$ and $G(\eta)$ 
in the most general ortho-toric metric given by \eqref{24},
\begin{equation}\label{410}
F(z) ~= ~G(z) ~= ~1 - z^2, 
\end{equation}
we get the following flat metric on $R^4$, which we will refer to 
as the \emph{ACG-quadratic} metric:
\begin{equation}\label{411}
ds^2 = (\xi - \eta) \left[\frac{d\xi^2}{1-\xi^2} + \frac{d\eta^2}{\eta^2-1} \right]+\frac{1}{\xi - \eta} \left[ (1- \xi^2) (dt +\eta \,dz)^2 + ( \eta^2 -1)(dt + \xi \,dz)^2\right].
\end{equation}
Requiring that the metric \eqref{411} be positive definite 
constrains the ranges of the $\xi$-$\eta$ co-ordinates 
to\footnote{We have made the choice $\xi \ge \eta$; the other 
choice $\xi \le \eta$ just interchanges $\xi$ and $\eta$.}.:
\begin{equation}\label{412}
-1 \leq \xi \leq 1, \qquad - \infty < \eta \leq -1,
\end{equation}
which is exactly the same region in the $\xi-\eta$ plane as the 
one occupied by the C-metric co-ordinates, see figure \ref{fig1}. 
We can work out the explicit co-ordinate transformations between 
the polar $\lbrace r_1, \phi, r_2, \psi \rbrace$ co-ordinates and 
the $ \lbrace \xi, t, \eta, z \rbrace$ co-ordinates.
\begin{eqnarray}\label{413}
r_1 = \sqrt{-2(1+\xi \,\eta + \xi + \eta)}, \qquad  \phi = \frac{t+z}{2} , \nonumber \\  
r_2 = \sqrt{2(1+\xi \,\eta - \xi - \eta)}, \qquad \psi = \frac{t-z}{2} . 
\end{eqnarray}
The inverse of the above are:
\begin{eqnarray}\label{414}
\xi = \frac{\Xi - r_1^2 - r_2^2}{8}, \qquad t = \phi + \psi, \nonumber \\
\eta = -\frac{\Xi + r_1^2 + r_2^2}{8}, \qquad z = \phi - \psi.
\end{eqnarray}
where 
\begin{equation}\label{415}
\Xi =  \sqrt{(8 + r_1^2 + r_2^2)^2 - 32~ r_2^2}.
\end{equation}

\subsection{C-metric co-ordinates $\quad \leftrightarrows \quad $ Darboux co-ordinates}

From \eqref{412} and \eqref{47}, it is clear that both the 
metrics \eqref{46} and \eqref{411} are defined in the same region 
(figure \ref{fig1}) of the $x$-$y/\xi$-$\eta$ plane. Furthermore, 
the two metrics \eqref{46} and \eqref{411} are isometric to each 
other, which can be checked using the following explicit 
co-ordinate transformations between the $\lbrace \xi, t, \eta, z 
\rbrace$ co-ordinates and the $\lbrace x, \phi, y, \psi \rbrace$ 
co-ordinates, which can be worked out from \eqref{49} and 
\eqref{413}:
\begin{eqnarray}\label{416}
x = \frac{ 4 ( \xi + \eta )+1}{\Lambda}, \qquad \phi = \frac{t+z}{2} \nonumber \\
y = \frac{ 4 ( \xi + \eta )-1}{\Lambda}, \qquad \psi = \frac{t-z}{2}, 
\end{eqnarray}
where
\begin{equation}\label{417}
\Lambda = \sqrt{1 - 8 ( 1 + \xi \,\eta) + 16 ( \xi + \eta )^2}.
\end{equation}
The inverse of the above are:
\begin{eqnarray}\label{418}
\xi = \frac{ x+y + \Upsilon}{8 \,( x-y)}, \qquad t = \phi +\psi, \nonumber \\
\eta = \frac{x+y - \Upsilon}{8 \,(x-y)}, \qquad z = \phi - \psi,
\end{eqnarray}
where
\begin{equation}\label{419}
\Upsilon = \sqrt{ 4 + 28\, ( 1 - x \,y) + 7 \,(x - y)^2}.
\end{equation}
We will need the above in section five to co-ordinate transform 
various asymptotically flat solutions known in C-metric 
co-ordinates to the Darboux co-ordinates.

Both the C-metric and the Darboux co-ordinates share an exchange 
symmetry $x \leftrightarrow y$, $\xi \leftrightarrow \eta$, which 
simplifies computations and allows for compact expressions. We 
want to think of the Darboux co-ordinates as ``ring-like'' 
co-ordinates because they are so closely related to the C-metric 
``ring co-ordinates.'' A compelling reason to consider flat space 
in these Darboux co-ordinates is that we can simply deform the 
metric and the complex structure while keeping the K\"ahler form 
fixed to arrive at the base-space of $AdS_5$, as we will show in 
the next section.

\section{$AdS_5$ in ring-like co-ordinates}

\subsection{The ACG-cubic Metric}

Choosing the same cubic polynomial for $F(\xi)$ and $G(\eta)$ in 
the most general ortho-toric metric given by \eqref{24},
\begin{equation}\label{301}
F(z) ~= ~G(z) ~= ~(1 - z^2)\,(1+ a\,p - z\,p) , 
\end{equation}
with $a > (p-1)/p$ and $p > 0$ real constants, we get the 
following 
K\"ahler metric which we will refer to as the \emph{ACG-cubic} metric:
\begin{eqnarray}\label{31}
ds^2_{ACG-cubic} & = &  (\xi - \eta) \Bigl[ \frac{d\xi^2}{(1 - \xi^2)\,(1+ a\,p - \xi\,p)} - \frac{d\eta^2}{(1 - \eta^2)\,(1+ a\,p - \eta\,p)} \Bigr] \\  & +  &\frac{1}{\xi - \eta} \Bigl[ (1 - \xi^2)\,(1+ a\,p - \xi\,p) \,  (dt + \eta \, dz)^2 - (1 - \eta^2)\,(1+ a\,p - \eta\,p)\, ( dt + \xi \,dz)^2 \Bigr]. \nonumber
\end{eqnarray}

We find that for the co-ordinate ranges \eqref{412}, the 
ACG-cubic metric is positive definite. Thus the ACG-cubic metric 
is defined in the same region of the $\xi-\eta$ plane as the 
ACG-quadratic metric (figure \ref{fig1}.) Furthermore, setting $p 
\rightarrow 0$ gives us back the ACG-quadratic metric. Some 
properties of the ACG-cubic metric are:
\begin{itemize}
\item[\textbf{(i)}] The ACG-cubic metric is a K\"ahler-Einsten 
metric. \newline 
When one chooses a cubic polynomial as above, we 
are assured (by a consequence of Propositions 4 and 11 of 
\cite{6}) to get a K\"ahler-Einstein metric. One can compute and 
verify that the ACG-cubic metric is a K\"ahler-Einstein metric 
with $- 6 \,p$ for it's Ricci scalar,
\begin{equation}\label{32}
R_{\mu \nu} = - \frac{ 3\, p}{2} g_{\mu \nu}.
\end{equation}

\item[\textbf{(ii)}] The ACG-cubic metric has \emph{constant 
holomorphic sectional curvature}. \newline
A K\"ahler metric with a constant holomorphic sectional curvature 
is the K\"ahler geometry equivalent of a maximally symmetric 
metric in Riemannian geometry. The Riemann tensor can be 
decomposed into three pieces: the fully traceless Weyl tensor 
($W$), a partially traceless part that can be written in terms of 
the metric and the traceless Ricci tensor (call it $P$)  and a 
trace that involves only the Ricci scalar (call it $S$). A metric 
with \emph{constant sectional curvature} corresponds to having $W 
= P = 0$, which leads to the usual expression for a maximally 
symmetric metric:
\begin{equation}\label{33}
R_{ i j k l } = S_{ i j k l} = \frac{ R}{d ( d -1)} \left(g_{i k }\, g_{ j l } - g_{i l } \,g_{ j k }  \right).
\end{equation} 
In K\"ahler geometry, employing the extra symmetries of the 
Riemann tensor, one can similarly decompose again into three 
pieces (section 2.63 of \cite{12}): a fully traceless part 
(called $B_0$ in \cite{12}), a partially traceless part that can 
be written in terms of the K\"ahler form and the traceless Ricci 
form (call it $P'$) and a piece that involves only the scalar 
curvature (call it $S'$.) A metric with constant holomorphic 
sectional curvature corresponds to having $B_0 = P' = 0$, which 
leads to the following expression for the Riemann tensor:
\begin{equation}\label{34}
R_{ i j k l } = S'_{ i j k l} = \frac{ R}{4 \frac{d}{2} ( \frac{d}{2} +1)} \left(g_{i k }\, g_{ j l } - g_{i l }\, g_{ j k }  +  2\, \Omega_{i j}\, \Omega_{k l} - \Omega_{i l}\, \Omega_{j k} + \Omega_{j l}\, \Omega_{i k} \right).
\end{equation}
One can see that a metric with constant holomorphic sectional 
curvature is not maximally symmetric. The ACG-cubic metric 
\eqref{31} satisfies equation \eqref{34}, which we can verify 
explicitly.\footnote{ We were informed by Harvey Reall that he 
and his collaborators were able to arrive at a closely related 
version of the ACG-cubic metric by imposing constant holomorphic 
sectional curvature on a certain class of Plebanski-Demianski 
metrics. }

\item[\textbf{(iii)}] The ACG-cubic metric is holomorphically 
isometric to the Bergmann metric.
\newline
The Bergmann metric is a K\"ahler metric defined for the unit 
ball $\mathbf{B}^4 \subset \mathbf{C}^2$ (section 3.2 of 
\cite{Gauntlett:2003fk}) in complex co-ordinates
\begin{equation}\label{35}
z^1 = r \, \cos {\textstyle\frac{\theta}{2}} ~e^{\frac{i ( \phi + \psi )}{2} }, \qquad z^2 = r\, \sin {\textstyle\frac{\theta}{2}} ~e^{\frac{i ( \phi - \psi )}{2} }. 
\end{equation}
using the K\"ahler potential:
\begin{equation}\label{36}
K = - \frac{2}{p} \log (1 - \vert z^1 \vert^2 - \vert z^2 \vert^2),
\end{equation} 
where the real co-ordinates in \eqref{35} are the radius $r$ and 
the Euler angles. The explict form of the metric is:
\begin{equation}\label{3601}
ds^2_{Bergmann} = \frac4p \left[ \frac{dr^2}{(1 - r^2)^2} + \frac{r^2}{4 ( 1 - r^2)} (d\theta^2 + \sin {\textstyle\theta}^2 d\phi^2) + \frac{r^2}{4 ( 1 - r^2)^2}  (d\psi + \cos \theta \,d\phi)^2 \right]
\end{equation} The Bergmann metric has a constant holomorphic 
sectional curvature. There is a theorem in K\"ahler geometry that 
says that \emph{ any two simply-connected complete K\"ahler 
manifolds with constant holomorphic sectional curvature are 
holomorphically isometric to each other} (see for example 
\cite{13}, page 179.) We can invoke this theorem to conclude that 
the ACG-cubic metric is holomorphically isometric to the Bergmann 
metric.

\item[\textbf{(iv)}] The ACG-cubic metric has a constant 
Kretschmann scalar, i.e.
\begin{equation}\label{361}
R_{ i j k l } R^{ i j k l } \vert_{ACG-cubic} = 12 \,p^2.
\end{equation}
This is an indication that the metric is a homogeneous 
metric.\footnote{ We thank Toby Wiseman for suggesting this 
test for a homogeneous metric.} Given that the Bergmann metric is 
known to be a homogeneous metric, it being the metric on the coset 
space $\frac{SU(2,1)}{U(2)}$, this adds further weight to the 
conclusion that the ACG-cubic metric and the Bergmann metric are 
holomorphically isometric to each other. \end{itemize}

So far we have given various arguments towards concluding that the ACG-cubic metric is isometric to the Bergmann metric. In appendix A, we provide an explicit co-ordinate transformation between the polar co-ordinates and the ring-like Darboux co-ordinates.

\subsection{$AdS_5$} 

With a base space that is K\"ahler-Einstein, we can get a 
solution to the Gauntlett-Gutowski theory i.e. a solution in the 
time-like class to minimal gauged supergravity by choosing $f =1$ 
and the one-form $\omega$ as (section 3.2 of \cite{Gauntlett:2003fk},)
\begin{eqnarray}\label{362}
\omega_{AdS_5} &=& \sqrt{p}\;[ (\xi + \eta)~ dt + ( 1 + \xi \,\eta )~ dz ] \nonumber \\
&=& \sqrt{p}\; \Omega^{(1)}
\end{eqnarray}
where $\Omega^{(1)}$ is the one-form that is natural to the 
K\"ahler structure given in \eqref{281}. We can check by an 
explicit computation that the metric,
\begin{equation}\label{37}
ds^2 = - [ d\tau +  \omega_{AdS_5} ]^2 ~+ ~ds^2_{ACG-cubic}.
\end{equation}
is a maximally symmetric metric with Ricci scalar $-5\, p$. To 
confirm that this is indeed $AdS_5$ space, we can invoke the 
following theorem proved by Gutowski and Reall in \cite{Gutowski:2004ez}: 
\emph{the only maximally symmetric solution in the time-like 
class to minimal gauged supergravity is $AdS_5$ and the base 
space is locally isometric to the Bergmann manifold.}

To reiterate, in this section, we have obtained the Bergmann 
metric and $AdS_5$ in ring-like co-ordinates in the same 
co-ordinate region as the susic flat ring.

\section{The K\"ahler structure of susic flat solutions}

In this section, we will first co-ordinate transform known 
$U(1)^2$ invariant asymptotically flat susic solutions to the 
ring-like Darboux co-ordinates.

For the supersymmetric flat-ring solution \cite{Elvang:2004rt}, the $f$ and 
$\omega$ are given in C-metric co-ordinates by
\begin{equation}\label{52}
\frac{1}{f} = 1+\frac{Q-q^2}{2}  (x-y)-\frac{q^2}{4} \left(x^2-y^2\right)
\end{equation}
and $\omega = \omega_\phi ~d\phi + \omega_\psi ~d\psi$, with
\begin{eqnarray}
\omega_\phi = - \frac{q}{8} (1-x^2) [ 3 Q - q^2 ( 3 + x + y ) ] \label{53} \\
\omega_\psi = \frac{3 \,q}{2}  (1+y) + \frac{q}{8} ( 1-y^2) [ 3Q - q^2 ( 3 + x + y ) ]. \label{54}
\end{eqnarray}
where $Q$ is the charge of the ring and $q$ it's dipole charge. 
Using the co-ordinate transformations \eqref{416}, the flat-ring 
solution takes the following form in the ring-like Darboux 
co-ordinates.
\begin{equation}\label{55}
\frac{1}{f} = 1 + \frac{Q-q^2}{ \Lambda} - 4 \,q^2\, \frac{\xi + \eta}{\Lambda^2}
\end{equation}
where $\Lambda$ is given in \eqref{417}. The one-form $\omega$ is 
now $\omega = \omega_t ~dt + \omega_z ~dz$, with
\begin{eqnarray}
\omega_t &=& ~~\frac{3\, q}{4}+\frac{3 \,q }{4\, \Lambda} (4 \xi +4 \eta -1)+~\frac{3 \,q \left(Q-q^2\right) }{\Lambda^2} (\xi +\eta )~ -\frac{8 \,q^3 }{\Lambda^3}(\xi +\eta )^2 \label{56} \\
\omega_z &=& -\frac{3 \,q}{4}-\frac{3\, q }{4\, \Lambda} (4 \xi +4 \eta -1)+\frac{3\, q \left(Q-q^2\right) }{\Lambda^2}(1+ \xi \eta ) -\frac{8 \,q^3 }{\Lambda^3}(\xi +\eta ) (1+ \xi \,\eta ) .\label{57}
\end{eqnarray}
It is possible to describe a supersymmetric black hole in the 
background of the supersymmetric black-ring by a simple 
modification of the $\frac{1}{f}$ of the solution 
\cite{Gauntlett:2004wh,Gauntlett:2004qy,Bena:2004de,Bena:2005zy}. The black hole by itself has a $SU(2) \times 
U(1)$ isometry and doesn't disturb the $U(1)^2$ isometry of the 
ring on its own. The modification involves adding the following 
harmonic function, harmonic with respect to the Laplacian on 
$R^4$,
\begin{equation}\label{58}
\frac{1}{f} = 1- \frac{Q_{BH}}{16} \left( \frac{x-y}{x+y}\right)+\frac{Q-q^2}{2}  (x-y)-\frac{q^2}{4} \left(x^2-y^2\right),
\end{equation}
where $Q_{BH}$ is the charge of the black hole. The one-form 
$\omega$ for this system is given by
\begin{eqnarray}
\omega_\phi = - \frac{q}{8} (1-x^2) \left[ 3 Q - q^2 ( 3 + x + y ) -\frac{9\, Q_{BH}}{4} \frac{1}{x+y} + 2 K \frac{1}{(x+y)^2} \right] \label{59} \\
\omega_\psi = \frac{3\, q}{2}  (1+y) + \frac{q}{8} ( 1-y^2)  \left[ 3 Q - q^2 ( 3 + x + y ) -\frac{9 \,Q_{BH}}{4} \frac{1}{x+y} + 2 K \frac{1}{(x+y)^2} \right], \label{510}
\end{eqnarray}
where $K$ is a constant proportional to the rotation of the black 
hole. Co-ordinate transforming to the ring-like Darboux 
co-ordinates, we have,
\begin{equation}\label{511}
\frac{1}{f} = 1 - \frac{Q_{BH}}{64} \frac{1}{\xi+\eta} + \frac{Q-q^2}{ \Lambda} - 4 \,q^2\, \frac{\xi + \eta}{\Lambda^2}
\end{equation}
and the one-form $\omega$ is:
\begin{eqnarray}
\omega_t &=& \omega_{t(0)} - ~\frac{9 \,q\, Q_{BH}}{32 \,\Lambda} + \frac{ K \,q}{32\, ( \xi + \eta )}  \label{512} \\
\omega_z &=& \omega_{z(0)} - \frac{9 \,q\, Q_{BH}}{32\, \Lambda} \frac{1 + \xi \eta}{ \xi + \eta} +  \frac{ K\, q ( 1 + \xi \eta) }{32\, ( \xi + \eta )^2} \label{513}
\end{eqnarray}
where $\omega_{t(0)} $ and $\omega_{z(0)} $ are the expressions 
for pure black ring (i.e. without the black hole) as given in 
formulae \eqref{56} and \eqref{57}.

We first note that the $\omega$ for the supersymmetric flat ring 
\eqref{56}, \eqref{57} can be written in the following suggestive 
manner\footnote{It is perhaps not irrelevant to note that the 
$\omega$ for the non-supersymmetric flat ring of \cite{Emparan:2001wn} also is 
of the form \eqref{515} with $\kappa (\xi,\eta) =0$; of course, 
the stationary form of the metric is not as simple as 
\eqref{01}.}:
\begin{eqnarray}\label{515}
\omega &=& \rho(\xi,\eta) ~\frac{ dt - dz }{2} + \frac{\kappa(\xi, \eta)}{\Lambda^2} ~[ (\xi + \eta)~ dt + ( 1 + \xi \eta )~ dz ] \nonumber \\
 &=& \rho(\xi,\eta) ~d\psi + \frac{\kappa(\xi, \eta)}{\Lambda^2} ~\Omega^{(1)},
\end{eqnarray}
where $\Omega^{(1)}$ is the one-form natural to the K\"ahler 
structure given in \eqref{281}. For the flat ring, the functions 
$\rho$ and $\kappa$ are $\rho_0$ and $\kappa_0$ given by,
\begin{eqnarray}
\rho_0(\xi,\eta) =\frac{3 q}{2} \left( 1 + \frac{4 \xi + 4 \eta -1}{\Lambda} \right) \label{516} \\
\kappa_0 (\xi, \eta) = 3 q ( Q - q^2 ) - 8\, q^3 \frac{\xi +\eta}{\Lambda} . \label{517}
\end{eqnarray}
The $\omega$ for the black hole in the background of the flat 
ring also takes the above form \eqref{515}, for which the 
functions $\rho$ and $\kappa$
\begin{eqnarray}
\rho(\xi,\eta)_{BH \, in\, a\, Ring} &=& \rho_0(\xi,\eta) \label{518} \\
\kappa(\xi,\eta)_{BH \, in\, a\, Ring} &=& \kappa_0(\xi,\eta) - \frac{9 \,q \,Q_{BH}}{32} \frac{\Lambda}{ \xi + \eta} + \frac{K q}{32} \left( \frac{ \Lambda} {\xi + \eta} \right)^2. \label{519} 
\end{eqnarray}

This form for $\omega$ \eqref{515}, \eqref{022} seems to encode 
in it the fact that there is a ring (through the $d\,\psi$ term), 
the underlying K\"ahler structure (through the term 
$\Omega^{(1)}$) and the $U(1)^2$ invariance (because all known 
$U(1)^2$ invariant solutions take this form.)

\section{An Ansatz for a supersymmetric $AdS$ ring?}

\begin{figure}[t]
\centering
\includegraphics[scale=0.5]{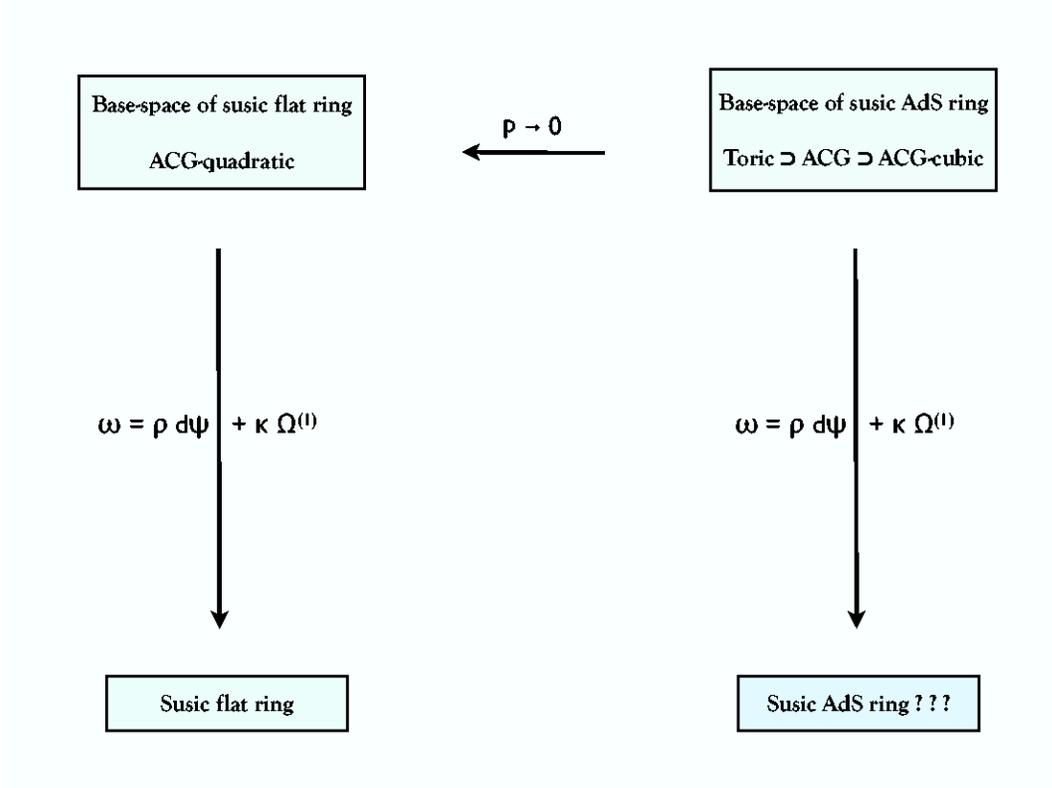}
\caption{A schematic representation of our approach to obtaining 
a supersymmetric $AdS$ ring.}
\label{fig2}
\end{figure}

Our approach to a susic $AdS$ ring is summarized 
schematically in figure \ref{fig2}.

\subsection{An Ansatz for the $AdS$ ring K\"ahler metric}

As summarized in the box below, we have many classes of $U(1)^2$ invariant 
K\"ahler metrics from which to pick an Ansatz for the $AdS$ ring.
$$
\boxed{
\ \phantom{\Big|}\textrm{Weakly self-dual ortho-toric} \ \subset \  
\textrm{ACG 
ortho-toric}\ \subset \ \textrm{Toric K\"ahler metrics.}\ 
}
$$
The first class, i.e. weakly-self-dual ortho-toric, has the most 
explicit classification; it is completely determined by the 
quartic polynomials \eqref{28}. To decide if we can make an 
Ansatz for the $AdS$ ring with a weakly self-dual ortho-toric 
K\"ahler metric, we checked if the Gutowski-Reall black hole 
\cite{Gutowski:2004ez,Gutowski:2004yv} K\"ahler metric is weakly self-dual and found that 
it \emph{isn't}. The details of this can be found in the 
appendix. Hence, we rule out a weakly-self-dual Ansatz.
 
The second class i.e. the ACG metrics \eqref{24} admit a less 
explicit classification; there are two arbitrary functions in the 
game. The functions are functions of one variable, which is a 
simplifying feature, giving total derivatives rather than partial 
derivatives for the connections and curvatures.  The third class 
i.e the toric K\"ahler metrics \eqref{21} admit an even less 
explicit classification; the metric involves functions of two 
variables. One could simplify things a bit by working with a 
single function of two variables, the symplectic potential. It 
remains to be seen which of the two classes of metrics will 
contain the $AdS$ ring. We will report further progress in this 
direction in future work \cite{14}.  Both classes of metrics 
include the two limiting cases viz. the asymptotic ACG-cubic 
metric and the $p \rightarrow 0$ limit, i.e. the ACG-quadratic 
metric.

\subsection{An Ansatz for $\omega$}

The form \eqref{515} for the one-form $\omega$ seems to capture 
some crucial features that we desire for a supersymmetric $AdS$ 
ring, viz. K\"ahler structure, $U(1)^2$ invariance etc.  Since 
the asymptotic $AdS_5$ metric and the full $AdS$ ring K\"ahler 
metric Ansatz share the same K\"ahler structure (i.e. both 
$\Omega$ and $\Omega^{(1)}$), we can happily make the same Ansatz 
\eqref{515} for the $ \omega$ of the supersymmetric $AdS$ ring.  
We also have the following information for the Ansatz functions 
$\rho$ and $\kappa$. At asymptotic infinity, $\rho \rightarrow 0, 
\kappa \rightarrow \sqrt{p} ~\Lambda^2$.  And as $p \rightarrow 
0$, $\rho \rightarrow \rho_0, \kappa \rightarrow \kappa_0$. The 
observation that $\kappa(\xi, \eta)$ is a function of the 
combination $\frac{\xi + \eta}{\Lambda}$ should also be useful.

We now need to plug these Ans\"atze into the Gauntlett-Gutowski 
theory. The Gauntlett-Gutowski theory of supersymmetric solutions 
works somewhat differently to the theory of supersymmetric 
solutions to gauged supergravity. First, the $\frac{1}{f}$ of the 
solution gets completely determined by the Ricci scalar of the 
base-space K\"ahler metric and second, the $G^+$ of the solution 
gets completely determined by the traceless Ricci form of the 
base-space K\"ahler metric. These two facts constrains our 
Ans\"atze above. Finally the $G^-$ of the solution is determined 
by a first-order p.d.e which itself is determined fully by the 
base-space K\"ahler metric.\footnote{We would also need to take 
care of the addition to the Gauntlet-Gutowski theory given in 
\cite{Figueras:2006xx}, which ensures that the $d\omega$ thus determined is 
indeed closed.} We will leave this for future work \cite{14}.

\section*{Conclusion}

Starting from the requirements of $U(1)^2$ isometry and 
a K\"ahler base, which are natural to supersymmetric AdS ring 
solutions, enabled us to get some insight into the supersymmetric 
solutions. 
Coordinates (related 
to the) moment maps of the two $U(1)$'s turn out to be 
``ring-like.'' We have described $AdS_5$ and other solutions in
these ring-like 
co-ordinates and have given an Ans\"atz for 
the supersymmetric $AdS$ ring solution and a strategy 
for obtaining it (figure \ref{fig2}.)  
Since \cite{Kunduri:2006uh}, \cite{Kunduri:2007qy} casts doubt on the 
existence of supersymmetric $AdS$ rings, our Ansatz provides a
context for addressing this question further.
Recently a non-supersymmetric black ring solution in de-Sitter 
space was constructed \cite{Chu:2006pf}. This is encouraging because it 
suggests that non-supersymmetric black rings can exist in 
$AdS$.
The toric almost-K\"ahler\footnote{ An 
almost K\"ahler manifold is not a complex manifold and hence not 
K\"ahler.} metrics of \eqref{21} provide a metric Ansatz 
to start with for non-supersymmetric $AdS$ rings.

\section*{Acknowledgments}

We would like to thank Justin David, Edward Teo, Martin O' 
Loughlin, Harvey Reall, Boris Pioline, Koushik Ray and Toby 
Wiseman for discussions.

\section*{Appendix A.}

\textbf{The Fubini-Study metric as an ACG-cubic metric:}
\newline 
The ACG-cubic metric \eqref{31} has two parameters, $a$ and $p$ while the Bergmann metric has only one parameter $p$. To conclude that the two metrics are isometric, we need to show that $a$ is a co-ordinate artifact or even better display an explicit co-ordinate transformation that settles this issue. We do this in the context of  Fubini-Study metric on $\mathbf{CP}^2$, which is the compact version of the Bergmann metric. It can be simply obtained by replacing $(1 - r^2) \rightarrow (1+ r^2)$ in \eqref{3601}. More precisely, it is the K\"ahler metric obtained from the K\"ahler potential,
\begin{equation}\label{A1}
K =  \frac{2}{p} \log (1 + \vert z^1 \vert^2 + \vert z^2 \vert^2),
\end{equation} 
which takes the explicit form,
\begin{equation}\label{A2}
ds^2_{FS} = \frac4p \left[ \frac{dr^2}{(1 + r^2)^2} + \frac{r^2}{4 ( 1 + r^2)} (d\theta^2 + \sin {\textstyle\theta}^2 d\phi^2) + \frac{r^2}{4 ( 1 + r^2)^2}  (d\psi + \cos \theta \,d\phi)^2 \right],
\end{equation}
in the complex structure
\begin{equation}\label{A3}
z^1 = r \, \cos {\textstyle\frac{\theta}{2}} ~e^{\frac{i ( \phi + \psi )}{2} }, \qquad z^2 = r\, \sin {\textstyle\frac{\theta}{2}} ~e^{\frac{i ( \phi - \psi )}{2} }. 
\end{equation}
The Fubini-Study metric is also a K\"ahler-Einstein metric  with constant holomorphic sectional curvature. In this appendix, we will show that the Fubini-Study metric can be described as an ACG-cubic metric. First consider the following general ACG-cubic metric,
\begin{equation}\label{A4}
F(x) = G(x) = - p \,( x - b_1) (x - b_2) ( x - b_3).
\end{equation}
Either by direct computation or using the results of \cite{6}, one can ascertain that this is a K\"ahler-Einstein metric of Ricci scalar $ 6\  p$. We can also directly compute and verify that  the above ACG-cubic metric has constant holomorphic sectional curvature.  Under the co-ordinate transformation
\begin{eqnarray}\label{A5}
x_1 = \xi + \eta, &\qquad& x_2 = \xi \, \eta, \\ \label{A6} 
\xi = \frac{1}{2} \left(x_1+\sqrt{x_1^2-4 x_2}\right), &\qquad& \eta = \frac{1}{2} \left(x_1- \sqrt{x_1^2-4 x_2}\right)
\end{eqnarray}
to the moment map co-ordinates, the ACG-cubic metric takes a certain form which we won't reproduce here. But more importantly, this metric can be derived from the following symplectic potential,
\begin{align}\label{A7}
G_S(x_1,x_2) = - \frac{ x_2 - b_1 x_1 + b_1^2 }{p\, (b_1 - b_2) (b_3 - b_1)} \log \left[- \frac{ 2 (x_2 - b_1 x_1 + b_1^2 )}{p\,(b_1 - b_2) (b_3 - b_1)}\right] \nonumber \\ - \frac{x_2 - b_2 x_1 + b_2^2 }{p\, (b_2 - b_3) (b_1 - b_2)} \log \left[- \frac{2 ( x_2 - b_2 x_1 + b_2^2 )}{p \, (b_2 - b_3) (b_1 - b_2)}\right] \nonumber \\  - \frac{x_2 - b_3 x_1 + b_3^2 }{p\, (b_3 - b_1) (b_2 - b_3)} \log \left[- \frac{2 ( x_2 - b_3 x_1 + b_3^2 )}{p\,(b_3 - b_1) (b_2 - b_3)} \right]
\end{align}
using the formula \eqref{21}. We will not go into the details of how we obtained \eqref{A7} from the ACG data \eqref{A4}. For the complete theory of symplectic potentials for the ACG metrics and every other consideration in this appendix, we refer the reader to \cite{Balasubramanian:2007dv}. From the work of Guillemin \cite{8}, it is known that the above symplectic potential \eqref{A7} encodes in it the co-ordinate singularities of a toric manifold whose moment polytope is the convex set enclosed within the lines, 
\begin{equation}\label{A8}
l_i \equiv x_2 - b_i \, x_1 + b _i^2 = 0, \qquad i = 1,2,3.
\end{equation}
In the $x_1 - x_2$ plane, $l_i = 0$ is a line with slope $b_i$ and $x_2$-intercept equalling $ - b_i^2$. The intersection of the lines are:
\begin{equation}\label{A9}
l_1 \cap l_2: (b_1 + b_2, b_1\, b_2), \qquad l_2 \cap l_3: (b_2 + b_3, b_2\, b_3), \qquad l_3 \cap l_1: (b_3 + b_1, b_3\, b_1). 
\end{equation}
We thus have that the moment polytope of the ACG-cubic metric \eqref{A4} is a triangle with vertices \eqref{A9} and edges \eqref{A8}. 

At this stage, we have used up all the local properties of the Fubini-Study metric, namely K\"ahler-Einstein and constant holomorphic sectional curvature, and we still have three undetermined constants in the ACG-cubic metric \eqref{A3} viz. $b_1, b_2, b_3$.  The Fubini-Study metric \eqref{A2} has no free parameters apart from the Ricci-scalar ($\sim p$). We will have to use the global features of the Fubini-Study metric and $\mathbf{CP}^2$ to fix the parameters $b_i$. Global considerations specify the size and shape of the triangle which in turn fixed the $b_i$'s. Before we do this, we give the co-ordinate transformation between the Darboux co-ordinates and the polar co-ordinates:
\begin{eqnarray}\label{A90}
r^2 &=& \frac{ b_1 \left(b_2-b_3\right)-b_2 b_3+(\xi +\eta ) \, b_3-\xi  \eta}{b_3^2-(\xi +\eta ) b_3+\xi  \eta },  \nonumber \\ \tan^2 \theta/2 &=& \frac{(b_2^2-(\xi +\eta ) \, b_2+\xi  \eta )(b_3-b_1)}{(b_1^2-(\xi +\eta )\,  b_1+\xi  \eta) ( b_2-b_3)}, \nonumber \\ \phi &=& -\frac{p}{2} \left(b_1-b_2\right) \left(t+b_3 z\right), \nonumber  \\ \psi &=& \frac{p}{2} \left[( 2 b_3 - b_1 - b_2 ) t-( 2 b_1 b_2 - b_1 b_3 - b_2 b_3 ) z\right], 
\end{eqnarray}
and the inverse, 
\begin{eqnarray} \label{A10}
\xi + \eta &=& \frac{ 2 ( b_1 + b_2 )  + ( b_1 + b_2 + 2 b_3 ) r^2 - ( b_1 - b_2 ) r^2 \cos \theta }{2(1 + r^2)}, \nonumber \\  \xi \, \eta &=& \frac{ 2  b_1  b_2   + ( b_1 + b_2 ) b_3 \,  r^2 - ( b_1 - b_2 )b_3 \, r^2 \cos \theta }{2(1 + r^2)}, \nonumber \\  t &=& \frac{( 2 b_1 b_2 - b_2 b_3  - b_3 b_1 ) \phi + ( b_2 b_3 - b_3 b_1 ) \psi}{p\, \left(b_1-b_2\right) \left(b_2-b_3\right) \left(b_3-b_1\right)},  \nonumber  \\ z &=& \frac{( 2 b_3 - b_1 - b_2 ) \phi + ( b_1 -  b_2 ) \psi}{p\, \left(b_1-b_2\right) \left(b_2-b_3\right) \left(b_3-b_1\right)} . 
 \end{eqnarray}
 To get the above, one first goes from symplectic co-ordinates to complex co-ordinates by a Legendre transform \cite{8}, \cite{9}. The Legendre transform also provides the K\"ahler potential. Then one matches with \eqref{A1} and \eqref{A3}. It is clear, especially from \eqref{A90}, that the parameters $b_i$ appear only in the co-ordinate transformation, and hence do not carry any local co-ordinate invariant information. 
 \newline\textbf{Global Considerations:}
\newline
Now, we can fix the parameters $b_i$ by the following two facts that carry global information of the Fubini-Study metric and $\mathbf{CP}^2$.
\begin{itemize}

\item[\textbf{(i)}] {The moment polytope of $\mathbf{CP}^2$ is a right-angled isosceles triangle.}

\item[\textbf{(ii)}] {The volume of $\mathbf{CP}^2$ is fixed once one specifies the Ricci-scalar. For \eqref{A2}, it is $\frac{8\,\pi^2}{p^2}$}

\end{itemize}
The easiest way of seeing \textbf{(i)} is to note that in the GLSM description of $\mathbf{CP}^2$, there is one D-term constraint, $| \phi_1 |^2  + | \phi_2 |^2 | + | \phi_3 |^2 = \frac{2}{p}$, which is solved by 
\begin{equation}\label{A11}
| \phi_1 |^2 = x_1, \qquad | \phi_2 |^2 = x_2, \qquad | \phi_3 |^2 = \frac{2}{p} - x_1 - x_2.
\end{equation}
The moment polytope is then given by the region 
\begin{equation}\label{A12}
| \phi_1 |^2  > 0 , \qquad | \phi_2 |^2  > 0 , \qquad | \phi_3 |^2  > 0,
\end{equation}
which is clearly a right-angled isosceles triangle of area $\frac{2}{p^2}$. 

Let us choose the right angle to be at the intersection of $l_1 =  0$ and $ l_2 = 0$. We then have the following (two independent) constraints:
\begin{equation}\label{A13}
b_1\, b_2 = -1, \qquad \frac{b_1 - b_3}{1 + b_1 \, b_3} = 1, \qquad  \frac{b_3 - b_2}{1 + b_3 \, b_2} = 1.
\end{equation}
This allows us to fix two of the parameters in terms of one free parameter, say $b_3$:
\begin{equation}\label{A14}
b_1 = \frac{1+b_3}{1-b_3}, \qquad b_2 = - \frac{1-b_3}{1+b_3}.
\end{equation}
Now we will impose the volume constraint. In the symplectic co-ordinates, by virtue of the fact that the four-metric is of a block-diagonal form \eqref{21} with two $2\times 2$ matrices which are inverses of each other, the determinant of the metric is just $1$. Hence, the volume of the manifold is just the product of the Euclidean  volumes of the moment polytope and the angular torus. From \eqref{A9} and from \eqref{A14}, we can compute the Euclidean volume of the moment polytope in terms of $b_3$. To compute the volume of the angular torus, we need the co-ordinate transformation \eqref{A10}. Using the ranges of the angles $0 < \phi < 2 \pi, ~  0 < \psi < 4 \pi$ and the second two equations in \eqref{A10}, we can compute the Euclidean volume of the angular torus. Putting everything together, we have 
\begin{equation}\label{A15}
\textrm{Vol}_{ACG-cubic} = \frac{4 \pi^2}{p^2} \frac{(2 - b_3) (1 + 2 b_3)}{1+b_3^2}.
\end{equation}
Now, requiring $\textrm{Vol}_{ACG-cubic} = \textrm{Vol}_{\mathbf{CP}^2} = \frac{8 \pi^2}{p^2}$, we get a quadratic equation for $b_3$ with two roots $b_3 = 0, \frac34$. The two roots suggest that we have two different ways of realizing the Fubini-Study metric as an ortho-toric ACG metric. One of the solutions, $b_3 =0$ gives us co-ordinate singularities which are ``ring-like.'' Using \eqref{A14}, we then have,
\begin{equation}\label{A16}
b_1 = 1, \qquad b_2 = -1, \qquad b_3 = 0,
\end{equation}
so that the polynomials occuring in the ACG-cubic metric are 
\begin{equation}\label{A17}
F(z)~ = ~G(z)~ =~  (1 - z^2)\, ( p \, z) .
\end{equation}
The explicit co-ordinate transformation is now much simpler: 
\begin{eqnarray}\label{A18}
 r^2 = - \frac{ 1 + \xi \eta}{\xi \eta}, \qquad  \tan^2 \theta/2 = \frac{ 1 + \xi \eta +  ( \xi + \eta) }{ 1 + \xi \eta - ( \xi + \eta) }, \quad  \phi = - p\,  t, \quad \psi &=& p \, z; \nonumber \\ \xi + \eta = - \frac{ r^2 \cos \theta}{1+r^2}, \quad \xi \, \eta = - \frac{1}{1+ r^2}, \quad   t = - \frac{ \phi}{p}, \quad z =  \frac{\psi}{p}.
\end{eqnarray}  

In this appendix, we have been able to show the explicit isometry between the Fubini-Study metric and a compact ACG-cubic metric by deriving an explicit co-ordinate transformation between the polar co-ordinates of the form and the ``ring-like'' co-ordinates of the latter.  A similar co-ordinate transformation relates the Bergmann metric to the ACG-cubic metric; the co-efficient of the cubic has the opposite sign reflecting the sign of the scalar curvature. However, one free parameter remains as Bergmann space is non-compact.
 
\section*{Appendix B.}
\textbf{The Gutowski-Reall black hole K\"ahler metric and weak self-duality:}
\newline 
In the recent past, many researchers have constructed various 
supersymmetric $AdS_5$ black holes \cite{Gutowski:2004ez,Gutowski:2004yv,Kunduri:2006ek, Chong:2005da, Chong:2005hr}, but 
we will only need to consider the first ones and the simplest of 
them, the Gutowski-Reall black hole. The K\"ahler metric of the 
Gutowski-Reall black holes \cite{Gutowski:2004ez,Gutowski:2004yv} is
\begin{equation}\label{29}
ds^2 = dr^2 + a^2 ((\sigma_L^1)^2 + (\sigma_L^2)^2)+ ( 2 a a' )^2 (\sigma_L^3)^2,
\end{equation}
where $a(r)$ is 
\begin{equation}\label{291}
a(r) =  {\textstyle\sqrt{\frac{\alpha}{4}+\frac{1}{p}}} \,\sinh {\textstyle\frac{\sqrt{p} \,r}{2}},
\end{equation}
with $\alpha$ is a constant and $\sigma_L^i$ are the 
right-invariant one-forms on $SU(2)$, which can be expressed in 
terms of the Euler angles $(\theta, \phi, \psi)$ as
\begin{eqnarray}\label{210}
\sigma_L^1 &=& \sin \phi ~ d\theta - \cos \phi ~ \sin \theta ~ d\psi, \nonumber \\
\sigma_L^2 &=& \cos \phi ~ d\theta + \sin \phi ~ \sin \theta ~ d\psi, \nonumber \\
\sigma_L^3 &=&  d\phi +  \cos  \theta ~ d\psi.
\end{eqnarray}
This metric \eqref{29} is K\"ahler with the K\"ahler form
\begin{equation}\label{211}
\Omega = ~ d ( a^2 \sigma_L^3 )~  = ~ 2 a a' ~ dr \wedge \sigma_L^3 ~-~ a^2 ~\sigma_L^1 \wedge \sigma_L^2
\end{equation}
In the orientation with $ \textrm{Vol} = 2 a^3 a' \sin  \theta \,dr \wedge 
d\theta \wedge d\phi \wedge d\psi$, the K\"ahler form \eqref{211} 
is self-dual.  We can check for the weak-self-duality property of 
a given K\"ahler metric in atleast two ways \cite{6}.

\noindent
\textbf{(i)} The defining property of a weakly self-dual K\"ahler 
metric is that the anti-self-dual Weyl tensor is harmonic, i.e. 
$\delta W^- = 0$ where $\delta$ acts on $W^-$ as on a two-form 
with values in anti-self-dual two-forms. On the other hand, the 
Weyl tensor on a Riemannian four-manifold is linked with the 
Cotton-York tensor, $ \delta W^\pm ~=~C^\pm $. Hence, vanishing 
of the anti-self-dual Cotton-York tensor implies the weak 
self-duality of a given K\"ahler metric (see definition 2 in 
\cite{6}.)

The Cotton-York tensor is defined in \cite{6}, $C_{X,Y} (Z) := ( 
\nabla_Y h)(X,Z) - ( \nabla_X h) (Y,Z) $, where $h$ is the 
normalized Ricci tensor given by $h_{\mu \nu} = \frac{1}{2} 
R_{\mu \nu} - \frac{R}{12} g_{\mu \nu}.$ Define it's components 
by $C_{X,Y}(Z):=C_{k i j} Z^k X^i Y^j$ and we get
\begin{equation}\label{212}
C_{k i j} = \frac{1}{2} \left( \nabla_j R_{i k} -  \nabla_i R_{j k} \right) - \frac{1}{12} \left(  g_{i k} \nabla_j R  - g_{j k} \nabla_i R  \right).
\end{equation}
The indices $i,j$ are anti-symmetric and one is supposed to think 
of the Cotton-York tensor as a co-vector-(the index $k$)-valued 
two-form, $C_r^{(2)}, C_\theta^{(2)}, C_\phi^{(2)}, 
C_\psi^{(2)}$. Plugging in \eqref{29} into \eqref{212}, we get
\begin{eqnarray}
C_r^{(2)} &=& 0, \nonumber \\
C_\theta^{(2)} &=& -{\textstyle\frac{\alpha \, \sqrt{p^3}}{48}} \coth {\textstyle\frac{\sqrt{p} \,r}{2}} ~ dr \wedge d\theta + {\textstyle\frac{\alpha\,p^3}{256} \left( 1 + \frac{\alpha \, p }{4} \right)} \cosh^2 {\textstyle\frac{\sqrt{p} \,r}{2}} \sin \theta ~ d\phi \wedge d\psi, \nonumber \\
C_\phi^{(2)} &=& {\textstyle\frac{\alpha\,\sqrt{p^7}}{384} \left( 1 + \frac{\alpha \, p }{4} \right)}   \cosh^2 {\textstyle\frac{\sqrt{p} \,r}{2}} ~\coth {\textstyle\frac{\sqrt{p} \,r}{2}}~ dr \wedge \sigma_L^3 + {\textstyle\frac{\alpha\,\sqrt{p^3}}{128} \left( 1 + \frac{\alpha \, p }{4} \right)} \cosh^2 {\textstyle\frac{\sqrt{p} \,r}{2}} ~\sigma_L^1 \wedge \sigma_L^2, \nonumber \\
C_\psi^{(2)} &=& \ldots \label{213}
\end{eqnarray} 
Computing the anti-self-dual parts of the above two-forms gives us,
\begin{eqnarray}
C_r^{-} &=& 0, \nonumber \\
C_\theta^{-} &=& -{\textstyle \frac{\alpha\, \sqrt{p^3}}{24}} \coth {\textstyle\frac{\sqrt{p} \,r}{2}} ~ dr \wedge d\theta + {\textstyle\frac{\alpha\,p^3}{384} \left( 1 + \frac{\alpha \, p }{4} \right)}  \cosh^2 {\textstyle\frac{\sqrt{p} \,r}{2}} \sin \theta ~ d\phi \wedge d\psi, \nonumber \\
C_\phi^{-} &=& {\textstyle\frac{\alpha\,\sqrt{p^7}}{192} \left( 1 + \frac{\alpha \, p }{4} \right)}  \cosh^2 {\textstyle\frac{\sqrt{p} \,r}{2}} ~\coth {\textstyle\frac{\sqrt{p} \,r}{2}}~ dr \wedge \sigma_L^3 + {\textstyle\frac{\alpha\, p^3}{192} \left( 1 + \frac{\alpha \, p }{4} \right)}  \cosh^2 {\textstyle\frac{\sqrt{p} \,r}{2}} ~\sigma_L^1 \wedge \sigma_L^2, \nonumber \\
C_\psi^{-} &=& \ldots \label{214}
\end{eqnarray} 
We thus find that the anti-self-dual part of the Cotton-York 
tensor is non-vanishing for the black hole K\"ahler metric, thus 
making it \emph{not} weakly self-dual. Note that for $\alpha = 
0$, the K\"ahler metric is weakly self-dual, which is as it 
should be because it corresponds to the Bergmann K\"ahler metric.

\noindent
\textbf{(ii)} On a complex two-dimensional K\"ahler manifold $(g, 
J, \Omega_J)$ with orientation chosen so that the K\"ahler form, 
$\Omega_J$, is self-dual, a property of the traceless Ricci form, 
$\rho_0$, is that it is anti-self-dual. Any anti-self-dual 
two-form can always be written as a functional multiple of an 
anti-self-dual K\"ahler form of some other (needn't be 
integrable) almost-complex structure, $\rho_0 = \lambda~ 
\Omega_I$, where $(g, I, \Omega_I)$ is an almost-hermitian 
structure. One of the many defining properties of a weakly 
self-dual K\"ahler metric is that $( \frac{g}{\lambda^2}, I, 
\frac{\Omega_I}{\lambda^2})$ is K\"ahler (see lemma 2 and lemma 4 
of \cite{6}.)

For the K\"ahler metric \eqref{29}, the traceless Ricci form is 
\begin{equation}\label{215}
\rho_0 =   b ~ dr \wedge \sigma_L^3 + \frac{ b\, a}{2 a'} ~ \sigma_L^1 \wedge \sigma_L^2,
\end{equation}
where $b(r)$ is
\begin{equation}\label{216}
b(r) = - \frac{ a'''  a^2 + 3 a'' a' a - 4 a'^3 + a'}{a}.
\end{equation}
This $\rho_0$ can be written as $\lambda~ \Omega_I$ with
\begin{equation}\label{217}
\Omega_I = ~ 2 a a' ~ dr \wedge \sigma_L^3 ~+~ a^2 ~\sigma_L^1 \wedge \sigma_L^2
\end{equation}
and 
\begin{equation}\label{218}
\lambda(r) = \frac{b}{2 a a'}.
\end{equation}
It is easy to see that $\Omega_I$ is anti-self-dual and one can 
construct the almost complex structure $I \sim g^{-1}~\Omega_I$ 
and verify $I^2 = - Id$.  For the K\"ahler metric \eqref{29} to 
be weakly self-dual,
\begin{equation}\label{2181}
d~  \left(~ \frac{\Omega_I}{\lambda^2}~\right)   ~ = ~ 0,
\end{equation}
which amounts to the following fourth order non-linear o.d.e:
\begin{equation}\label{219}
\left( \frac{a'^2 a^6}{b^2} \right)' + \frac{2 a'^3 a^3}{b^2} = 0.
\end{equation}
But, the black hole K\"ahler metric \eqref{291} does \emph{not} 
satisfy the above o.d.e, hence \emph{not} weakly self-dual.

\end{document}